\documentclass[a4paper,10pt,onecolumn,preprint,sort&compress]{elsarticle}
\usepackage[margin=3.15cm]{geometry}

\makeatletter
\def\ps@pprintTitle{%
 \let\@oddhead\@empty
 \let\@evenhead\@empty
 \def\@oddfoot{\centerline{\thepage}}%
 \let\@evenfoot\@oddfoot}
\makeatother

\usepackage{epsfig,graphicx,amsmath,amssymb}
\usepackage{mathtools,cuted}
\usepackage{multirow}
\usepackage[english]{babel}
\usepackage{fancyhdr}
\usepackage{physics}
\usepackage{listings}
\usepackage{mathrsfs}
\usepackage{slashed}
\usepackage[colorinlistoftodos]{todonotes}
\usepackage[utf8x]{inputenc}
\usepackage{fancyhdr}
\usepackage{array}
\usepackage{subfig}
\usepackage{graphicx}
\usepackage{afterpage}
\usepackage{adjustbox}

\usepackage{lineno,hyperref}
\modulolinenumbers[5]

\begin{document}

\begin{frontmatter}

\title{$\boldsymbol{\pi^0$-$\eta$-$\eta^{\prime}}$ \textbf{mixing from} $\boldsymbol{V\!\rightarrow\!P\gamma}$ \textbf{and} $\boldsymbol{P\!\rightarrow\!V\gamma}$ \textbf{decays}}


\author[mymainaddress,mysecondaryaddress]{Rafel Escribano}
\ead{rescriba@ifae.es}

\author[mymainaddress]{Emilio Royo\corref{correspondingauthor}}
\cortext[correspondingauthor]{Corresponding author.}
\ead{emilio.royocarratala@e-campus.uab.cat}

\address[mymainaddress]{Grup de Física Teòrica, Departament de Física, Universitat Autònoma de Barcelona, E-08193 Bellaterra (Barcelona), Spain}
\address[mysecondaryaddress]{Institut de Física d’Altes Energies (IFAE), The Barcelona Institute of Science and Technology, Campus UAB, E-08193 Bellaterra (Barcelona), Spain}

\begin{abstract}
An enhanced phenomenological model that includes isospin-symmetry breaking is presented in this letter.
The model is then used in a number of statistical fits to the most recent experimental data for the radiative transitions
$V\!P\gamma$ ($V=\rho$, $K^*$, $\omega$, $\phi$ and $P=\pi$, $K$, $\eta$, $\eta^{\prime}$)
and estimations for the mixing angles amongst the three pseudoscalar states with vanishing third-component of isospin are obtained.
The quality of the performed fits is good, e.g.~$\chi^2_{\textrm{min}}/\textrm{d.o.f} = 1.9$.
The current experimental uncertainties allow for isospin-symmetry violations with a confidence level of approximately $2.5\sigma$.
\end{abstract}


\end{frontmatter}




\section{\label{intro}Introduction}

The flavour $SU(3)$ symmetry is broken by the strange quark being significantly heavier than the up and down quarks
\cite{Bramon:1997va,Feldmann:1999uf,Bickert:2016fgy}. 
As a result of this, the physical states $\eta$ and $\eta^{\prime}$ become a mixture of the pure octet $\ket{\eta_8}$ and singlet $\ket{\eta_0}$ mathematical states.
Through an orthogonal transformation with mixing angle $\theta_P$, the mass eigenstates $\ket{\eta}$ and $\ket{\eta^{\prime}}$
can be expressed as a linear combination of $\ket{\eta_8}$ and $\ket{\eta_0}$ \cite{Bramon:1997va,Bickert:2016fgy},
\begin{equation}
\begin{aligned}
\ket{\eta}=\cos{\theta_P}\ket{\eta_8}-\sin{\theta_P}\ket{\eta_0}\ ,\\
\ket{\eta^{\prime}}=\sin{\theta_P}\ket{\eta_8}+\cos{\theta_P}\ket{\eta_0}\ ,
\end{aligned}
\label{eq100}
\end{equation}
with $\ket{\eta_8} \!=\! \frac{1}{\sqrt{6}}\ket{u\bar{u} \!+\! d\bar{d} \!-\! 2s\bar{s}}$ and $\ket{\eta_0} \!=\! \frac{1}{\sqrt{3}}\ket{u\bar{u} \!+\! d\bar{d} \!+\! s\bar{s}}$.
Another commonly used basis for the description of the $\eta$-$\eta^{\prime}$ mixing is the quark-flavour basis,
which becomes exact in the limit $m_s \rightarrow \infty$ \cite{Feldmann:1998su},
\begin{equation}
\begin{aligned}
\ket{\eta}=\cos{\phi_P}\ket{\eta_{\textrm{NS}}}-\sin{\phi_P}\ket{\eta_{\textrm{S}}}\ ,\\
\ket{\eta^{\prime}}=\sin{\phi_P}\ket{\eta_{\textrm{NS}}}+\cos{\phi_P}\ket{\eta_{\textrm{S}}}\ , 
\end{aligned}
\label{eq101}
\end{equation}
where $\ket{\eta_{\textrm{NS}}} = \frac{1}{\sqrt{2}}\ket{u\bar{u} + d\bar{d}}$ and $\ket{\eta_{\textrm{S}}} = \ket{s\bar{s}}$.
The mixing angles $\theta_P$ and $\phi_P$ are related by $\theta_P = \phi_P - \arctan{\sqrt{2}} \simeq \phi_P - 54.7^\circ$.

The mixing of the $\eta$ and $\eta^{\prime}$ mesons is heavily influenced by the $U(1)_A$ anomaly of QCD \cite{Escribano:2005qq},
which induces a significant amount of mixing in the $\eta$-$\eta^{\prime}$ sector \cite{Feldmann:1999uf}. 
The $U(1)_A$ anomaly forces the $\ket{\eta}$ and $\ket{\eta^{\prime}}$ mass eigenstates, which one would naively expect to be almost \textit{ideally mixed},
to be nearly flavour octet and singlet states. In addition, the $U(1)_A$ anomaly is responsible for the non-Goldstone nature of the singlet state, 
forcing it to be massive even in the chiral limit. 
As a result of the mixing, the $U(1)_A$ anomaly is transferred to both the $\eta$ and $\eta^{\prime}$ mesons \cite{Bickert:2016fgy}.

In the vector meson sector, where the spins of the quark-antiquark bound states are parallel, 
the mixing between the $\omega$ and $\phi$ mesons is usually described using the quark-flavour basis, 
as there is no anomaly affecting this sector \cite{Bramon:2000fr,Feldmann:1999uf}. 
Accordingly, the mixing angle $\phi_V$ is small (about $3^\circ$ to $4^\circ$), 
which is consistent with the OZI-rule and becomes rigorous in the limit $N_c \rightarrow \infty$ \cite{Feldmann:1999uf}.

Early phenomenological studies on the $\eta$-$\eta^{\prime}$ mixing used experimental data to perform statistical fits in terms of the mixing angles.
One significant contribution was made by Gilman et al.~in the late 1980s \cite{Gilman:1987ax}, 
which provided an estimation of $\theta_P \simeq -20^\circ$ after a complete review of the empirical data available at the time. 
Subsequently, Bramon et al., \cite{Bramon:1997mf} and \cite{Bramon:1997va}, 
introduced in their phenomenological model corrections due to non-ideal mixing in the vector meson nonet and obtained a somewhat less negative mixing angle, 
i.e.~$\theta_P = (-16.9\pm 1.7)^\circ$ and $\theta_P = (-15.5\pm 1.3)^\circ$, respectively, 
where the former was deduced from the rich set of $J/\psi$ decays into a vector and a pseudoscalar meson 
whilst the latter came from a thorough analysis of many different decay channels. In Ref.~\cite{Bramon:1997va}, 
the flavour $SU(3)$-breaking corrections were introduced in terms of constituent quark mass differences 
whilst mixing with other pseudoscalar states like glueballs was neglected. 
Benayoun et al.~proposed in Ref.~\cite{Benayoun:1999fv} an approach based on a hidden local symmetry model,
supplemented with nonet symmetry breaking in the pseudoscalar sector. 
This approach achieved good agreement with experimental data, with exception of the $K^{*\pm}$ radiative decays, 
and found a pseudoscalar mixing angle $\theta_P \simeq -11^\circ$, 
which is consistent with the quadratic Gell-Mann-Okubo mass formula but in conflict with chiral perturbation theory ($\chi$PT) expectations. 
A value of $\phi_V \simeq 3^\circ$ was also found.

In 2001, Bramon et al.~\cite{Bramon:2000fr} introduced an additional source of flavour $SU(3)$-symmetry breaking by 
including a quantum mechanical extension for the $V\!P\gamma$ radiative decays. 
The phenomenological model assumed isospin symmetry and the expectation that, even though gluon annihilation channels induce $\eta$-$\eta^{\prime}$ mixing, 
they play a negligible role in $V\!P\gamma$ transitions, respecting, therefore, the OZI-rule~\cite{Bramon:2000fr}. 
The $V\!P\gamma$ decay couplings were expressed in terms of the mixing angles and relative spatial wavefunction overlaps; 
then, using experimental estimations for the decay couplings, the best fit values for the free parameters of the model were obtained. 
The quality of their fits was very good (e.g.~$\chi^2_{\textrm{min}}/\textrm{d.o.f.} = 0.7$) 
and the estimations for the mixing angles were found to be $\phi_P = (37.7\pm 2.4)^\circ$ 
and $\phi_V = (3.4\pm 0.2)^\circ$ using the experimental data available at the time. 
An important conclusion that was drawn is that the $SU(3)$-breaking effects originated from flavour dependence through the 
relative spatial wavefunction overlaps cannot be neglected. 

Ball et al.~presented in Ref.~\cite{Ball:1995zv} (see also Ref.~\cite{Kiselev:1992ms})
a different approach by assuming that the meson decay constants follow the pattern of particle state mixing, 
connecting the short-distance properties of mesons, i.e.~decay constants, with long-distance phenomena, i.e.~mass eigenstates mixing \cite{Feldmann:1998su}.
In particular, the $V\!P\gamma$ radiative decays were directly linked to the anomaly of the $AVV$ triangle diagram,
and the $SU(3)$-breaking effects were introduced by means of leptonic decay constants. 
A fit using experimental data for several $V\!P\gamma$ decay channels enabled an estimation for $\theta_P$ between $-20^\circ$ and $-17^\circ$. 
This strategy and subsequent enhancements introduced by others have been ubiquitous in the literature 
(e.g.~\cite{Escribano:2005qq,Feldmann:1998su,Feldmann:1999uf,Escribano:2005vz,Escribano:1999nh,Feldmann:2002kz,Schechter:1992iz}). 
In this context, phenomenological studies have confirmed that a two mixing angle scheme is required to properly describe the experimental data in the octet-singlet basis 
\cite{Escribano:2005qq,Feldmann:1998vh,Escribano:2015nra,Escribano:2015yup,Feldmann:1997vc}, 
whilst a single mixing angle suffices to achieve good agreement in the quark-flavour basis 
\cite{Escribano:2005qq,Feldmann:1998vh,Feldmann:1998sh,Escribano:2013kba,Escribano:2015nra,Feldmann:1998yc}, 
which is supported by large-$N_c$ $\chi$PT \cite{Leutwyler:1997yr,Kaiser:2000gs} at next-to-leading order. 
This appears to indicate that the difference between the two mixing angles in the octet-singlet basis is produced by an $SU(3)$-breaking effect, 
whereas in the quark-flavour basis the difference comes from an OZI-rule violating effect \cite{Feldmann:1999uf,Escribano:2005qq}.
In addition, at lowest order in $\chi$PT, one only requires a single mixing angle, which endorses 
Eqs.~(\ref{eq100}) and (\ref{eq101}).

Using this approach, Feldmann et al.~\cite{Feldmann:1998vh} provided theoretical (to first order in flavour symmetry breaking) 
and phenomenological estimations for $\theta_P$ of $-12.3^\circ$ (no error provided) and $(-15.4\pm 1.0)^\circ$, respectively. 
Likewise, Escribano et al.~\cite{Escribano:2005qq} found phenomenological values for $\theta_P = (-14.3\pm 1.0)^\circ$ 
and $\phi_V = (4.1\pm 2.2)^\circ$ using one mixing angle in the quark-flavour basis. 
As well as this, Kroll obtained in Ref.~\cite{Kroll:2005sd} values for $\theta_P$ of $(-13.2\pm 2.2)^\circ$ and $(-13.5\pm 1.1)^\circ$,
 employing two different sources of empirical data available at the time, i.e.~the PDG 2004 and KLOE collaboration, respectively.

The gluonic content of the $\eta$ and $\eta^{\prime}$ wavefunctions was analysed using empirical data from $V\!P\gamma$ decays in 
Refs.~\cite{Escribano:2007cd,Thomas:2007uy}.
The model that was employed followed Ref.~\cite{Bramon:2000fr}. 
It was found that the gluonic content for the $\eta$ and $\eta^{\prime}$ wavefunctions is consistent with zero, using the most up-to-date data at the time. 
Furthermore, it was again emphasized the importance of the secondary source of flavour $SU(3)$-symmetry breaking to achieve good agreement with experimental data.

Feldmann et al.~discussed in Ref.~\cite{Feldmann:1998sh} the effects of isospin-symmetry breaking, which is induced by the mass difference between the $u$ and $d$ quarks, 
as well as QED effects, using the theoretical framework first presented in Ref.~\cite{Leutwyler:1996np}. 
Mathematically, they expressed the admixtures of the $\eta$ and $\eta^{\prime}$ to the physical $\pi^0$ as \cite{Feldmann:1998sh}
\begin{equation}
\ket{\pi^0}=\ket{\pi_3}+\epsilon\ket{\eta}+\epsilon^{\prime}\ket{\eta^{\prime}}\ ,
\label{eq113}
\end{equation}
where $\ket{\pi_3}$ denotes the $I_3 = 0$ state of the pseudoscalar isospin triplet. 
By assuming a mixing angle of $\phi = 39.3^\circ$ for the $\eta$-$\eta^{\prime}$ system,
they found through the diagonalisation of the associated mass matrix that the mixing between the $\pi^0$ and $\eta$ mesons was $\epsilon = 1.4\%$, 
whilst the $\pi^0$-$\eta^{\prime}$ mixing was $\epsilon^{\prime} = 0.37\%$ (no errors associated to these theoretical estimations were provided).

Kroll, as a continuation of the previous work, highlighted in Ref.~\cite{Kroll:2005sd} that isospin-symmetry breaking is of order $(m_d-m_u)/m_s$ 
due to the effect of the $U(1)_A$ anomaly, which is embodied in the divergence of the singlet axial-vector current \cite{Gross:1979ur,Gasser:1984gg}. 
As a result of the mixing, the $U(1)_A$ anomaly is transferred to the $\pi^0$, $\eta$ and $\eta^{\prime}$ physical states. 
A simple generalisation of the quark-flavour mixing scheme (e.g.~\cite{Feldmann:1998sh,Feldmann:1999uf,Feldmann:2002kz}) 
allowed him to write the following theoretical expressions for the mixing parameters $\epsilon$ and $\epsilon^{\prime}$ \cite{Kroll:2005sd},
\begin{equation}
\begin{aligned}
\epsilon(z)= \cos{\phi}\Bigg[\frac{1}{2}\frac{m_{dd}^2-m_{uu}^2}{m_{\eta}^2-m_{\pi^0}^2}+z\Bigg]\ ,\\
\epsilon^{\prime}(z)=\sin{\phi}\Bigg[\frac{1}{2}\frac{m_{dd}^2-m_{uu}^2}{m_{\eta^{\prime}}^2-m_{\pi^0}^2}+z\Bigg]\ ,
\end{aligned}
\label{eq114}
\end{equation}
where the parameter $z$ is the quotient of decay constants $z = (f_u - f_d)/(f_u + f_d)$ and the quark mass difference $m_{dd}^2 - m_{uu}^2$ 
was estimated from the $K^0$-$K^+$ mass difference. Assuming again a mixing angle in the $\eta$-$\eta^{\prime}$ sector of $\phi = 39.3^\circ$ 
and making use of the $f_u=f_d$ limit, he found the following numerical estimations for the mixing parameters $\epsilon$ and $\epsilon^{\prime}$,
\begin{equation}
\begin{aligned}
\hat{\epsilon}=\epsilon(z=0)=(1.7\pm 0.2) \% \ ,\\
\hat{\epsilon}^{\prime}=\epsilon^{\prime}(z=0)=(0.4\pm 0.1) \% \ .
\end{aligned}
\label{eq114b}
\end{equation}

Escribano et al.~analysed in Ref.~\cite{Escribano:2016ntp} the second-class current decays $\tau^- \rightarrow \pi^-\eta^{(\prime)}\nu_{\eta}$ 
and found estimations for the $\pi^0$-$\eta$ and $\pi^0$-$\eta^{\prime}$ mixing parameters from theory, 
making use of scalar and vector form factors at next-to-leading order in $\chi$PT. 
The analytic expressions that they found are consistent with those from Kroll shown in Eq.~(\ref{eq114}) up to high-order isospin corrections. 
The numerical estimations that they obtained are
\begin{equation}
\begin{aligned}
\epsilon_{\pi\eta}&=\textrm{c}\phi_{\eta\eta^{\prime}}\frac{m_{K^0}^2-m_{K^+}^2-m_{\pi^0}^2+m_{\pi^+}^2}
{m_{\eta}^2-m_{\pi^-}^2}\Bigg[1-\frac{m_{\eta}^2-m_{\pi^-}^2}{M_S^2} \Bigg]\\ 
&=(9.8\pm 0.3)\times 10^{−3}\ ,\\
\epsilon_{\pi\eta^{\prime}}&=\textrm{s}\phi_{\eta\eta^{\prime}}\frac{m_{K^0}^2-m_{K^+}^2-m_{\pi^0}^2+m_{\pi^+}^2}
{m_{\eta^{\prime}}^2-m_{\pi^-}^2}\Bigg[1-\frac{m_{\eta^{\prime}}^2-m_{\pi^-}^2}{M_S^2} \Bigg]\\
&=(2.5\pm 1.5)\times 10^{−4}\ .
\end{aligned}
\label{eq115}
\end{equation}
where $\textrm{c}\phi_{\eta\eta^{\prime}}$ and $\textrm{s}\phi_{\eta\eta^{\prime}}$ stand for $\cos{\phi_{\eta\eta^{\prime}}}$ and $\sin{\phi_{\eta\eta^{\prime}}}$; 
also, an $\eta$-$\eta^{\prime}$ mixing angle of $\phi_{\eta\eta^{\prime}} = (41.4\pm 0.5)^\circ$ was assumed, together with a scalar mass limit of $M_S=980$ MeV. 

It must be noted that Kroll's mixing parameters $\epsilon$ and $\epsilon^{\prime}$ in Ref.~\cite{Kroll:2005sd} (cf.~Eq.~(\ref{eq113})) 
were defined in the quark-flavour basis whilst Escribano et al.'s $\epsilon_{\pi\eta}$ and $\epsilon_{\pi\eta^{\prime}}$ in Ref.~\cite{Escribano:2016ntp} 
were defined making use of the octet-singlet basis. 
Despite this difference, it can be easily shown that, given that both authors used the same $SO(3)$ rotation matrix \textit{structure}, 
one can write $\epsilon=\epsilon_{\pi\eta}$ and $\epsilon^{\prime}=\epsilon_{\pi\eta^{\prime}}$, which are valid as first order approximations.



\section{\label{method}Methodology}

From the effective Lagrangian that is commonly used to describe $V\!P\gamma$ radiative decays, 
a set of expressions for the theoretical decay couplings is found in terms of the free parameters of the model. 
Next, using experimental data from Ref.~\cite{Tanabashi:2018oca}, 
the corresponding experimental decay couplings are calculated and, finally, an optimization fit can be performed.

In the framework of the conventional quark model, 
the flavour symmetry-breaking mechanism associated to differences in the effective magnetic moments of light and strange quarks in magnetic dipolar transitions 
is introduced via constituent quark mass differences. 
This is implemented by means of a multiplicative $SU(3)$-breaking term, i.e.~$1-s_e \equiv \overline{m}/m_s$, 
in the $s$-quark entry of the quark-charge matrix $Q$ \cite{Bramon:1997va}. 
A second source of flavour symmetry breaking, connected to the differences in the spatial extensions of the meson state wavefunctions, 
is also considered \cite{Bramon:2000fr}. 
This symmetry-breaking mechanism is introduced through additional multiplicative factors in the theoretical coupling constants, 
accounting for the corresponding relative wavefunction overlaps, and are left as free parameters in the fit. 

The isospin violation in the pseudoscalar sector is investigated in this framework. 
The mixing in this case requires an $SO(3)$ rotation matrix relating the $\pi^0$, $\eta$ and $\eta^{\prime}$ mass eigenstates to the $SU(3)$ mathematical states, 
with three mixing angles. 
Additional wavefunction overlap factors are introduced to the model and gluon annihilation channels, which might contribute to the mixing, 
are neglected\footnote{This is a necessary simplification 
to reduce the number of free parameters in the model; otherwise, the statistical fit would not be possible given the limited number of available decay channels.}.



\section{\label{sect3}The mixing of the $\eta$-$\eta^{\prime}$ revisited}

\begin{table}
\centering
\caption{Comparison between estimations for the seven free parameters from the model presented in Ref.~\cite{Bramon:2000fr}, 
using the PDG 2000 and the most up-to-date experimental data.}
{\def\arraystretch{1.2}\tabcolsep=45pt
\small
\begin{tabular}[c]{@{\hskip 0.05in}c @{\hskip 0.2in}c @{\hskip 0.2in}c @{\hskip 0.05in}}
\hline \hline
Parameter & Estimation from \cite{Bramon:2000fr} & Current Estimation\\ 
\hline \hline
g & $0.70 \pm 0.02 \ \textrm{GeV}^{-1}$ & $0.70 \pm 0.01 \ \textrm{GeV}^{-1}$ \\
$\frac{m_s}{\overline{m}}$ & $1.24 \pm 0.07$ & $1.17 \pm 0.06$ \\
$\phi_P$ & $(37.7 \pm 2.4)^\circ$ & $(41.4 \pm 0.5)^\circ$ \\
$\phi_V$ & $(3.4 \pm 0.2)^\circ$ & $(3.3 \pm 0.1)^\circ$ \\
$z_{\textrm{NS}}$ & $0.91 \pm 0.05$ & $0.84 \pm 0.02$ \\
$z_{\textrm{S}}$ & $0.89 \pm 0.07$ & $0.76 \pm 0.04$ \\
$z_{\textrm{K}}$ & $0.91 \pm 0.04$ & $0.89 \pm 0.03$ \\
\hline 
$\chi_{\textrm{min}}^2/\textrm{d.o.f.}$ & 0.7 & 4.6 \\
\hline \hline
\end{tabular}
}
\label{tb2}
\end{table}

The analysis carried out in Ref.~\cite{Bramon:2000fr} 
for the estimation of the mixing angle in the $\eta$-$\eta^{\prime}$ sector is reproduced in this section using the most up-to-date experimental data \cite{Tanabashi:2018oca}. 
The theoretical $V\!P\gamma$ decay couplings are confirmed to be those presented in Ref.~\cite{Bramon:2000fr}.
The relationship between the decay couplings and the decay widths is given by
\begin{equation}
\Gamma(V \rightarrow P\gamma) = \frac{1}{3}\frac{g_{V\!P\gamma}^2}{4\pi}\abs{\boldsymbol{p}_{\gamma}}^3 = \frac{1}{3}\Gamma(P \rightarrow V\gamma)\ ,
\label{eq121}
\end{equation}
where $\boldsymbol{p}_{\gamma}$ is the linear momentum of the outgoing photon. 
Using Eq.~(\ref{eq121}) together with the experimental data for the total decay widths, branching ratios and meson masses from Ref.~\cite{Tanabashi:2018oca}, 
one can obtain experimental values for the decay couplings.
From these and the corresponding theoretical counterparts, an optimisation fit can be performed. Making use of a standard minimisation software package, 
the optimal values for the seven free parameters of the model are presented in Table~\ref{tb2}.
One can see that the fitted values obtained in the present work are in good agreement with those found by Bramon et al.~in Ref.~\cite{Bramon:2000fr}. 
The current associated standard errors are smaller, which is due to the fact that the uncertainties associated to the experimental measurements have decreased over the years. 
The most recent empirical data seems to favour a somewhat bigger $\eta$-$\eta^{\prime}$ mixing angle $\phi_P$, 
which is consistent with other recent results (e.g.~Refs.~\cite{Escribano:2008rq,Ambrosino:2009sc,DiDonato:2011kr,Escribano:2013kba}). 
As well as this, the most up-to-date experimental data grants more relevance to the secondary source of flavour $SU(3)$-symmetry breaking, 
as the $z_{\textrm{NS}}$ and $z_{\textrm{S}}$ spatial wavefunction overlap factors are further from unity. 

That being said, the quality of the fit for the current estimations is poor with a $\chi^2_{\textrm{min}}/\textrm{d.o.f.} \simeq 23.1/5 \simeq 4.6$, while in Ref.~\cite{Bramon:2000fr},
using the data available at the time, the quality of the fit was excellent, i.e.~$\chi^2_{\textrm{min}}/\textrm{d.o.f.} = 0.7$. 
This, again, is connected to the improved quality of the most recent data \cite{Tanabashi:2018oca}. 
Based on this goodness-of-fit test, one ought to come to the conclusion that the current experimental data no longer supports the model presented in Ref.~\cite{Bramon:2000fr}.



\section{\label{sect4}Enhanced model for the $\pi^0$-$\eta$-$\eta^{\prime}$ mixing}

The phenomenological model presented above is enhanced in this section by incorporating isospin-breaking effects, 
enabling the investigation of the mixing phenomena between the $\pi^0$, $\eta$ and $\eta^{\prime}$ pseudoscalar mesons.
This improved model considers that the physical pseudoscalar mesons with vanishing third-component of isospin 
are an admixture of some pure mathematical states and the mixing is, thus, implemented by a three-dimensional rotation amongst them. 
In addition, the mechanisms of flavour $SU(3)$-symmetry breaking that have been discussed in section~\ref{method} are enhanced to account for violations of isospin. 
In the vector meson sector, a single mixing angle is still considered, as this sector is anomaly-free.

In order to find the theoretical decay couplings associated to the different $VP\gamma$ radiative transitions, 
one starts with the effective Lagrangian that is used to calculate amplitudes in $V \rightarrow P\gamma$ and $P \rightarrow V\gamma$ decay processes \cite{Bramon:1997va},
\begin{equation}
\mathscr{L}_{V\!P\gamma}=g_e\epsilon_{\mu \nu \alpha \beta}\partial^{\mu}A^{\nu}\textrm{Tr}[Q(\partial^{\alpha}V^{\beta}P+P\partial^{\alpha}V^{\beta})]\ ,
\label{eq150}
\end{equation}
where $g_e$ is a generic electromagnetic coupling constant, $\epsilon_{\mu \nu \alpha \beta}$ is the totally antisymmetric tensor, $A_{\mu}$ is the electromagnetic field, 
$V_{\mu}$ and $P$ are the matrices for the vector and pseudoscalar meson fields, respectively, and $Q$ is the quark-charge matrix $Q=diag\{2/3,-1/3,-1/3\}$ \cite{Bramon:1997va}. 

Next, the following $SO(3)$ rotation matrix correlating the pseudoscalar $I_3=0$ physical states with the pure quark-flavour basis states is selected
\begin{equation}
 \begin{pmatrix}
  \pi^0 \\
  \eta \\
  \eta^{\prime} \\
 \end{pmatrix}
 =
 \begin{pmatrix}
  1 & \epsilon_{12} & \epsilon_{13}  \\
  -\epsilon_{12}\textrm{c}\phi_{23} + \epsilon_{13}\textrm{s}\phi_{23} & \textrm{c}\phi_{23} & - \textrm{s}\phi_{23}  \\
 -\epsilon_{13}\textrm{c}\phi_{23} - \epsilon_{12}\textrm{s}\phi_{23} & \textrm{s}\phi_{23} & \textrm{c}\phi_{23}  \\
 \end{pmatrix} 
 \begin{pmatrix}
  \pi_3 \\
  \eta_{\textrm{NS}} \\
  \eta_{\textrm{S}} \\
 \end{pmatrix},
 \label{eq153}
\end{equation}
where $\epsilon_{12}$ and $\epsilon_{13}$ are first order approximations to the corresponding $\phi_{12}$ and $\phi_{13}$ mixing angles, 
as isospin-breaking corrections are small \cite{Leutwyler:1996np}. 
It must be stressed that the particular \textit{structure} that we have selected for the $SO(3)$ rotation matrix is down to the fact that it enables 
an enhanced \textit{resolution} against the statistical uncertainties associated to both mixing parameters $\epsilon_{12}$ and $\epsilon_{13}$ simultaneously, 
once the optimisation fits are performed\footnote{This point will become clearer later when the results are discussed.}.

The transformations that map Kroll's $\epsilon$ and $\epsilon^{\prime}$ in the quark-flavour basis (cf.~Eq.~(\ref{eq113}) 
and Ref.~\cite{Kroll:2005sd}) and Escribano et al.'s $\epsilon_{\pi\eta}$ and $\epsilon_{\pi\eta^{\prime}}$ in the octet-singlet basis (cf.~Ref.~\cite{Escribano:2016ntp})) 
to the $\epsilon_{12}$ and $\epsilon_{13}$ in the quark-flavour basis used in this letter (cf.~Eq.~(\ref{eq153})) 
are\footnote{Given that these are orthogonal transformations, to move from one definition to the other in the opposite direction, one only needs to multiply by the transposed matrices.}
\begin{equation}
 \begin{pmatrix}
  \epsilon_{12} \\ 
  \epsilon_{13}
 \end{pmatrix}
 = 
 \begin{pmatrix}
 \textrm{c}\phi_P & \textrm{s}\phi_P\\ 
 -\textrm{s}\phi_P & \textrm{c}\phi_P
 \end{pmatrix} 
 \begin{pmatrix}
  \epsilon \\ 
  \epsilon^{\prime}
 \end{pmatrix}\ ,
 \label{eq1532}
\end{equation}
and
\begin{equation}
\!
 \begin{pmatrix}
  \epsilon_{12} \\ 
  \epsilon_{13}
 \end{pmatrix}
 = \frac{1}{\sqrt{3}}
 \begin{pmatrix}
 \textrm{c}\theta_P-\sqrt{2}\ \textrm{s}\theta_P\!\!&\!\!\textrm{s}\theta_P+\sqrt{2}\ \textrm{c}\theta_P \\ 
 -\textrm{s}\theta_P-\sqrt{2}\ \textrm{c}\theta_P\!\!&\!\!\textrm{c}\theta_P-\sqrt{2}\ \textrm{s}\theta_P
 \end{pmatrix} 
 \begin{pmatrix}
  \epsilon_{\pi\eta} \\ 
  \epsilon_{\pi\eta^{\prime}}
 \end{pmatrix}.
 \label{eq1533}
\end{equation}

At this point, one can obtain the expressions for the theoretical decay couplings of the enhanced phenomenological model. 
These are
\begin{equation}
\begin{aligned}
g_{\rho^0\pi^0\gamma}			&= g\Big(\frac{1}{3} + \epsilon_{12}z_{\textrm{NS}}\Big)\ , \quad	 g_{\rho^+\pi^+\gamma} = g\frac{z_+}{3}\ ,\\
g_{\rho^0\eta\gamma}			&= g\Big[\Big(z_{\textrm{NS}} - \frac{\epsilon_{12}}{3}\Big)c\phi_{23} + \frac{\epsilon_{13}}{3}s\phi_{23}\Big]\ ,\\
g_{\omega\pi^0\gamma}			&= g\Big[\Big(1 + \frac{\epsilon_{12}}{3}z_{\textrm{NS}}\Big)c\phi_V + \frac{2}{3}z_{\textrm{S}}\frac{\overline{m}}{m_s}\epsilon_{13}s\phi_V\Big]\ ,\\
g_{\eta^{\prime}\rho^0\gamma} 		&= g\Big[\Big(z_{\textrm{NS}} - \frac{\epsilon_{12}}{3}\Big)s\phi_{23} - \frac{\epsilon_{13}}{3}c\phi_{23}\Big]\ ,\\
g_{\omega\eta\gamma} 			&= g\Bigg\{\Big[\Big(\frac{z_{\textrm{NS}}}{3} - \epsilon_{12}\Big)c\phi_{23} + \epsilon_{13}s\phi_{23}\Big]c\phi_V - \frac{2}{3}z_{\textrm{S}}
							\frac{\overline{m}}{m_s}s\phi_{23}s\phi_V\Bigg\}\ ,\\
g_{\eta^{\prime}\omega\gamma}	&= g\Bigg\{\Big[\Big(\frac{z_{\textrm{NS}}}{3} - \epsilon_{12}\Big)s\phi_{23} - \epsilon_{13}c\phi_{23}\Big]c\phi_V + \frac{2}{3}z_{\textrm{S}}
							\frac{\overline{m}}{m_s}c\phi_{23}s\phi_V\Bigg\}\ ,\\
g_{\phi\pi^0\gamma} 			&= g\Big[\Big(1 + \frac{\epsilon_{12}}{3}z_{\textrm{NS}}\Big)s\phi_V - \frac{2}{3}z_{\textrm{S}}\frac{\overline{m}}{m_s}\epsilon_{13}c\phi_V\Big]\ ,\\
g_{\phi\eta\gamma}				&= g\Bigg\{\Big[\Big(\frac{z_{\textrm{NS}}}{3} - \epsilon_{12}\Big)c\phi_{23} + \epsilon_{13}s\phi_{23}\Big]s\phi_V + \frac{2}{3}z_{\textrm{S}}
							\frac{\overline{m}}{m_s}s\phi_{23}c\phi_V\Bigg\}\ ,\\
g_{\phi\eta^{\prime}\gamma}		&= g\Bigg\{\Big[\Big(\frac{z_{\textrm{NS}}}{3} - \epsilon_{12}\Big)s\phi_{23} - \epsilon_{13}c\phi_{23}\Big]s\phi_V - \frac{2}{3}z_{\textrm{S}}
							\frac{\overline{m}}{m_s}c\phi_{23}c\phi_V\Bigg\}\ ,\\
g_{K^{*0}K^0\gamma}			&= -\frac{1}{3}g\Big(1 + \frac{\overline{m}}{m_s}\Big)z_{\textrm{K}^0}=
							-\frac{1}{3}g\Big(1 + z_{\textrm{S}}\frac{\overline{m}}{m_s}\Big)z_{\textrm{K}^0}^{\prime}\ ,\\
g_{K^{*+}K^+\gamma}			&= \frac{1}{3}g\Big(2 - \frac{\overline{m}}{m_s}\Big)z_{\textrm{K}^+}=
							\frac{1}{3}g\Big(2 - z_{\textrm{S}}\frac{\overline{m}}{m_s}\Big)z_{\textrm{K}^+}^{\prime}\ ,
\end{aligned}
\label{eq163}
\end{equation}
where the wavefunction overlap parameters have been redefined as relative overlap factors~\cite{Bramon:2000fr}: $z_{\textrm{NS}} \equiv Z_{\textrm{NS}}/Z_3$,  
$z_{\textrm{S}} \equiv Z_{\textrm{S}}/Z_3$, $z_+ \equiv Z_+/Z_3$, $z_{\textrm{K}^0} \equiv Z_{\textrm{K}^0}/Z_3$ and $z_{\textrm{K}^+} \equiv Z_{\textrm{K}^+}/Z_3$. 
The generic electromagnetic coupling constant $g_e$ in Eq.~(\ref{eq150}) has been replaced by $g = Z_3g_e$ on the right hand side equalities of Eq.~(\ref{eq163}). 
In some instances, the overlap factors in the strange sector have been redefined to 
$z_{\textrm{K}^0}^{\prime} = z_{\textrm{K}^0}(1+\overline{m}/m_s)/(1+z_{\textrm{S}}\overline{m}/m_s)$ and 
$ \ z_{\textrm{K}^+}^{\prime} = z_{\textrm{K}^+}(2-\overline{m}/m_s)/(2-z_{\textrm{S}}\overline{m}/m_s)$ in order to avoid redundant free parameters. 
It is worth highlighting that Eq.~(\ref{eq163}) reduces to the couplings shown in Ref.~\cite{Bramon:2000fr} in the good $SU(2)$ limit, as expected.

A fit of the theoretical decay couplings from Eq.~(\ref{eq163}) to the experimental data for ten free parameters provides the following estimations
\begin{equation}
\begin{aligned}
g &= 0.69 \pm 0.01\ \textrm{GeV}^{-1}\ , 			&	z_+ &= 0.95 \pm 0.05\ ,\\
\phi_{23} &= (41.5 \pm 0.5)^\circ\ , 				&	\phi_{V} &= (4.0 \pm 0.2)^\circ\ ,\\
\epsilon_{12} &= (2.3 \pm 1.0)\ \%\ , 				&	\epsilon_{13} &= (2.5 \pm 0.9)\ \%\ ,\\
z_{\textrm{NS}} &= 0.89 \pm 0.03\ , 				&	z_{\textrm{S}}\overline{m}/m_s &= 0.65 \pm 0.01\ ,\\
z_{\textrm{K}^0}^{\prime} &= 1.01 \pm 0.04\ , 		&	z_{\textrm{K}^+}^{\prime} &= 0.76 \pm 0.04\ .
\end{aligned}
\label{eq164}
\end{equation}
The quality of the fit is relatively good, with $\chi^2_{\textrm{min}}/\textrm{d.o.f.} \simeq 4.6/2 = 2.3$. 
The fitted values for the mixing angles $\phi_{23}$ and $\phi_V$ are in very good agreement with recent published results 
(e.g.~\cite{Escribano:2005qq,Escribano:2007cd,Ambrosino:2009sc}). The $g$ and $m_s/\overline{m}$ 
(see Eq.~(\ref{eq165}) below for an estimation of the latter) are also consistent with those from other studies but, as highlighted by Bramon et al.~in Ref.~\cite{Bramon:2000fr}, 
these parameters are largely dependent on the particular model used; hence, comparison provides limited value. 

An important point to notice from Eq.~(\ref{eq164}) is that the estimations for $\epsilon_{12}$ and $\epsilon_{13}$ 
are very small but not compatible with zero with a confidence level of $2.3\sigma$ and $2.8\sigma$, respectively, assuming a Gaussian distribution for the error. 
The $\epsilon_{12}$ and $\epsilon_{13}$ values from our fit can be translated to Kroll's and Escribano et al.'s definitions for their $SO(3)$ rotation matrix yielding 
$\epsilon=\epsilon_{\pi\eta} = (0.1 \pm 0.9)\ \%$ and $\epsilon^{\prime}=\epsilon_{\pi\eta^{\prime}} = (3.4 \pm 0.9)\ \%$. 
It can be observed that our mixing parameters $\epsilon$ and $\epsilon_{\pi\eta}$ are compatible with zero, whilst our parameters 
$\epsilon^{\prime}$ and $\epsilon_{\pi\eta^{\prime}}$ are not consistent with zero with a confidence level of $3.8\sigma$. 
Clearly, all mathematical representations for the physical states are equivalent; however, the specific rotation matrix selected in Eq.~(\ref{eq153}) 
enables the simultaneous ascertainment that \textit{both} parameters controlling the mixing in the $\pi^0$-$\eta$ and $\pi^0$-$\eta^{\prime}$ sectors are incompatible with zero.

In addition, it is worth noting from our results that the contribution to the physical state $\ket{\pi^0}$ from the mathematical state $\ket{\eta_8}$ is significantly smaller 
(in fact, consistent with zero) than that from the pure singlet state $\ket{\eta_0}$. 
This is an interesting result as one would naively expect the amount of mixing in the $\pi^0$-$\eta$ system to be larger than the one found in the $\pi^0$-$\eta^{\prime}$ sector, 
based on mass arguments. 
This can be explained, though, by the fact that the $U(1)_A$ anomaly mediates $\eta_0\leftrightarrow\pi_3$ transitions and, therefore, 
provides an additional contribution to the associated mixing. Note that Escribano et al.~\cite{Escribano:2016ntp} made use of the large-$N_c$ limit in their calculations, 
which effectively rids the theory of the chiral anomaly; hence, the effect mentioned above does not surface in their estimations for the mixing parameters. 
On the other hand, Kroll obtained in Ref.~\cite{Kroll:2005sd} first order theoretical results for the mixing parameters, neglecting, thus, any high-order symmetry breaking corrections; 
this is a sound approximation for the $\eta$-$\eta^{\prime}$ system but might potentially compromise the results for the $\pi^0$-$\eta$ and $\pi^0$-$\eta^{\prime}$ 
sectors where the mixing parameters are very small.

Another fit is carried out fixing $\epsilon_{12} = \epsilon_{13} = 0$ and leaving all the other parameters free. 
The quality of the fit is significantly decreased with $\chi^2_{\textrm{min}}/\textrm{d.o.f.} \simeq 21.3/4 \simeq 5.3$, 
highlighting the fact that a certain amount of mixing between the neutral $\pi^0$ with the $\eta$ and $\eta^{\prime}$ mesons different from zero is required to correctly describe the data.

Fixing the parameters $z_+ = 1$ and $z_{\textrm{K}^0} = z_{\textrm{K}^+}$, 
which accounts for \textit{turning off} the secondary mechanism of isospin-symmetry breaking, and performing a fit with all the other parameters left free, we find
\begin{equation}
\begin{gathered}
\begin{aligned}
g &= 0.69 \pm 0.01 \ \textrm{GeV}^{-1}\ , 	&	m_s/\overline{m} &= 1.17 \pm 0.06\ ,\\ 
\phi_{23} &= (41.5 \pm 0.5)^\circ\ , 	&	\phi_{V} &= (4.0 \pm 0.2)^\circ\ ,\\
\epsilon_{12} &= (2.4 \pm 1.0)\ \%\ , 		&	\epsilon_{13} &= (2.5 \pm 0.9)\ \%\ ,\\ 
z_{\textrm{NS}} &= 0.89 \pm 0.03\ , 		&	z_{\textrm{S}} &= 0.77 \pm 0.04\ ,\\
\end{aligned}
\\
z_{\textrm{K}} = 0.90 \pm 0.03\ ,	
\end{gathered}
\label{eq165}
\end{equation}
where the quality of the fit is better, i.e. $\chi^2_{\textrm{min}}/\textrm{d.o.f.} \ \simeq 5.6/3 \simeq 1.9$. 
The $z$'s in Eq.~(\ref{eq165}) are different from unity, signalling that the secondary mechanism of flavour $SU(3)$-symmetry breaking is still required for the correct description of the experimental data. 
This statement can be tested by performing a fit where all the $z$'s are fixed to one and it is found that the quality of the fit is substantially decreased, 
i.e.~$\chi^2_{\textrm{min}}/\textrm{d.o.f.} \simeq 41.8/6 \simeq 7.0$.

The estimates for $\epsilon_{12}$ and $\epsilon_{13}$ in Eq.~(\ref{eq165}) are, again, not compatible with zero with a confidence level of 
$2.4\sigma$ and $2.8\sigma$, respectively.
In general, the estimations from Eq.~(\ref{eq165}) are very approximate to the ones shown in Eq.~(\ref{eq164}).
It is interesting to see that reducing the number of free parameters in the last fit leads to a substantial increase in the quality of the fit. 
This is related to the fact that, despite the residual $\chi_{\textrm{min}}^2$ being smaller when ten free parameters are employed, 
this reduction does not compensate for the loss of one degree of freedom. 
Accordingly, it appears that the introduction of the secondary mechanism of isospin-symmetry breaking is not required to reproduce the experimental data. 
For this reason, the degrees of freedom $z_+$, $z_{\textrm{K}^0}$ and $z_{\textrm{K}^+}$ will be fixed to $z_+ = 1$ and $z_{\textrm{K}^0} = z_{\textrm{K}^+}$ for any subsequent fits.

Two more statistical fits using the estimated values for $\epsilon_{12}$ and $\epsilon_{13}$ from Kroll \cite{Kroll:2005sd} and Escribano et al.~\cite{Escribano:2016ntp} 
can be performed. 
Starting with Kroll's estimations $\epsilon_{12} = (1.6 \pm 0.2)\ \%$ and $\epsilon_{13} = (-0.8 \pm 0.1)\ \%$ we obtain
\begin{equation}
\begin{gathered}
\begin{aligned}
g &= 0.69 \pm 0.01 \ \textrm{GeV}^{-1}\ ,	&	m_s/\overline{m} &= 1.17 \pm 0.06\ ,\\
\phi_{23} & = (41.4 \pm 0.5)^\circ\ ,	&	 \phi_{V} &= (3.1 \pm 0.1)^\circ\ ,\\
z_{\textrm{NS}} &= 0.86 \pm 0.0\ ,		&	z_{\textrm{S}} &= 0.77 \pm 0.04\ ,\\
\end{aligned}
\\
z_{\textrm{K}} = 0.90 \pm 0.03\ ,
\end{gathered}
\label{eq166}
\end{equation}
where the quality of the fit is significantly poorer, i.e. $\chi^2_{\textrm{min}}/\textrm{d.o.f.}\simeq 22.0/5 = 4.4$.
Likewise, using Escribano et al.'s $\epsilon_{12} = (7.5 \pm 0.2) \times 10^{-3}$ and $\epsilon_{13} = (-6.3 \pm 0.2) \times 10^{-3}$ and performing the fit once more,
the following results are found
\begin{equation}
\begin{gathered}
\begin{aligned}
g &= 0.70 \pm 0.01 \ \textrm{GeV}^{-1}\ ,	&	m_s/\overline{m} = 1.17 \pm 0.06\ ,\\
\phi_{23} &= (41.4 \pm 0.5)^\circ\ , 	&	\phi_{V} = (3.2 \pm 0.1)^\circ\ ,\\
z_{\textrm{NS}} &= 0.85 \pm 0.02\ ,		&	z_{\textrm{S}} = 0.77 \pm 0.04\ ,\\
\end{aligned}
\\
z_{\textrm{K}} = 0.90 \pm 0.03\ ,
\end{gathered}
\label{eq167}
\end{equation}
where the quality of the fit is similar to the previous one, i.e.~$\chi^2_{\textrm{min}}/\textrm{d.o.f.} \simeq 24.0/5 = 4.8$. 
This shows that the theoretical estimations for the mixing parameters $\epsilon_{12}$ and $\epsilon_{13}$ provided by Kroll \cite{Kroll:2005sd} 
and Escribano et al.~\cite{Escribano:2016ntp} do not appear to agree with the most recent experimental data \cite{Tanabashi:2018oca}. 
It must be stressed, though, that the phenomenological model presented in this letter is based on the relatively simple standard quark model with a quantum mechanical extension, 
whilst Refs.~\cite{Kroll:2005sd} and \cite{Escribano:2016ntp} used more sophisticated theoretical approaches. 
Having said this, those estimations had limited numerical input from experiment due to their intrinsic theoretical nature.
\begin{table}[t]
\centering
\caption{Summary of fitted values for the Fit 1, Fit 2, Fit 3, Fit 4 and Fit 5, 
corresponding to Eqs.~(\ref{eq164}), (\ref{eq165}), (\ref{eq166}), (\ref{eq167}), and~(\ref{eq168}), respectively.}
{\def\arraystretch{1.2}\tabcolsep=45pt
\small
\begin{tabular}[c]{@{\hskip 0.05in}c @{\hskip 0.12in}c @{\hskip 0.12in}c @{\hskip 0.12in}c @{\hskip 0.12in}c @{\hskip 0.12in}c @{\hskip 0.05in}}
\hline \hline
Parameter & Fit 1 & Fit 2 & Fit 3 & Fit 4 & Fit 5 \\ 
\hline \hline
$g \ \textrm{(GeV}^{-1}\textrm{)}$ & $0.69 \pm 0.01$ & $0.69 \pm 0.01$ & $0.69 \pm 0.01$ & $0.70 \pm 0.01$ & $0.69 \pm 0.01$ \\
$\epsilon_{12}$ & $(2.3 \pm 1.0)\ \%$ & $(2.4 \pm 1.0)\ \%$ & - & - & $(2.4 \pm 1.0)\ \%$ \\
$\epsilon_{13}$ & $(2.5 \pm 0.9)\ \%$ & $(2.5 \pm 0.9)\ \%$ & - & - & $(2.5 \pm 0.9)\ \%$ \\
$\phi_{23} \ (^\circ)$ & $41.5 \pm 0.5$ & $41.5 \pm 0.05$ & $41.4 \pm 0.5$ & $41.4 \pm 0.5$ & $41.5 \pm 0.5$ \\
$\phi_V \ (^\circ)$ & $4.0 \pm 0.2$ & $4.0 \pm 0.2$ & $3.1 \pm 0.1$ & $3.2 \pm 0.1$ & $4.0 \pm 0.2$ \\
$m_s/\overline{m}$ & - & $1.17 \pm 0.06$ & $1.17 \pm 0.06$ & $1.17 \pm 0.06$ & - \\
$z_{\textrm{S}}\overline{m}/m_s$ & $0.65 \pm 0.01$ & - & - & - & $0.65 \pm 0.01$ \\
$z_{\textrm{NS}}$ & $0.89 \pm 0.03$ & $0.89 \pm 0.03$ & $0.86 \pm 0.02$ & $0.85 \pm 0.02$ & $0.89 \pm 0.03$ \\
$z_+$ & $0.95 \pm 0.05$ & - & - & -  & - \\
$z_{\textrm{S}}$ & - & $0.77 \pm 0.04$ & $0.77 \pm 0.04$ & $0.77 \pm 0.04$ & - \\
$z_{\textrm{K}}$ & - & $0.90 \pm 0.03$ & $0.90 \pm 0.03$ & $0.90 \pm 0.03$ & - \\
$z_{\textrm{K}^0}^{\prime}$ & $1.01 \pm 0.04$ & - & - & - & -\\
$z_{\textrm{K}^+}^{\prime}$ & $0.76 \pm 0.04$ & - & - & - & - \\
\hline 
$\chi_{\textrm{min}}^2/\textrm{d.o.f.}$ & 2.3 & 1.9 & 4.4 & 4.8 & 1.9 \\
\hline \hline
\end{tabular}
}
\label{tb3}
\end{table}

A final fit is carried out where the experimental points associated to the neutral and charged $K^* \rightarrow K\gamma$ transitions are not considered\footnote{Note that, 
traditionally, strange decay width measurements have suffered from larger uncertainties than the other radiative decays.}. 
Accordingly, the free parameters $z_{\textrm{K}}$, or $z_{\textrm{K}^0}^{\prime}$ and $z_{\textrm{K}^+}^{\prime}$, are not included in this fit, 
and the parameters $m_s/\overline{m}$ and $z_{\textrm{S}}$ are considered jointly again. 
The estimated values from the fit are
\begin{equation}
\begin{gathered}
\begin{aligned}
g &= 0.69 \pm 0.01 \ \textrm{GeV}^{-1} \ ,	&	z_{\textrm{S}}\overline{m}/m_s = 0.65 \pm 0.01 \ ,\\
\phi_{23} &= (41.5 \pm 0.5)^\circ \ ,	&	\phi_{V} = (4.0 \pm 0.2)^\circ \ ,\\
\epsilon_{12} &= (-2.4 \pm 1.0)\ \% \ ,	&	\epsilon_{13} = (-2.5 \pm 0.9)\ \% \ ,\\
\end{aligned}
\\
z_{\textrm{NS}} = 0.89 \pm 0.03 \ .
\end{gathered}
\label{eq168}
\end{equation}
The quality of the fit is good, $\chi^2_{\textrm{min}}/\textrm{d.o.f.}\simeq 5.6/3 \simeq 1.9$. 
The estimates for $\epsilon_{12}$ and $\epsilon_{13}$ are again incompatible with zero at a confidence level of $2.4\sigma$ and $2.8\sigma$, respectively.

A summary of all the fitted parameters is shown in Table~\ref{tb3}. 
The robustness of the fitted values for the parameters $g$, $\epsilon_{12}$, $\epsilon_{13}$, $\phi_{23}$ and $\phi_V$ across Fits 1, 2 and 5 is remarkable. 
In addition, the consistency of the $z$ parameters across all the fits is also very good.
As well as this, a comparison between the calculated decay widths and the experimental decay widths obtained directly from \cite{Tanabashi:2018oca} is presented in Table~\ref{tb4}.
The agreement is very good for the estimated values from $\Gamma_{\textrm{fit1}}$, $\Gamma_{\textrm{fit2}}$ and $\Gamma_{\textrm{fit5}}$. 
The decay width estimations $\Gamma_{\textrm{fit3}}$ and $\Gamma_{\textrm{fit4}}$ are not as good as the others, 
implying again that the experimental data seems to favour different values for $\epsilon_{12}$ and $\epsilon_{13}$ 
than those suggested by Kroll \cite{Kroll:2005sd} and Escribano et al.~\cite{Escribano:2016ntp}. 
\begin{table}[t]
\centering
\caption{Comparison between the experimental decay widths $\Gamma_{\textrm{exp}}$ for the various radiative decay channels and the 
$\Gamma_{\textrm{fit1}}$, $\Gamma_{\textrm{fit2}}$, $\Gamma_{\textrm{fit3}}$, $\Gamma_{\textrm{fit4}}$ and $\Gamma_{\textrm{fit5}}$ 
predictions from the enhanced model associated to the fit values from Eqs.~(\ref{eq164}), (\ref{eq165}), (\ref{eq166}), (\ref{eq167}), and~(\ref{eq168}), respectively.}
{\def\arraystretch{1.2}\tabcolsep=45pt
\small
\begin{tabular}[c]{@{\hskip 0.01in}c @{\hskip 0.03in}c @{\hskip 0.06in}c @{\hskip 0.06in}c @{\hskip 0.06in}c @{\hskip 0.06in}c @{\hskip 0.06in}c @{\hskip 0.01in}}
\hline \hline
Transition & $\Gamma_{\textrm{exp}}$ (keV) & $\Gamma_{\textrm{fit1}}$ (keV) & $\Gamma_{\textrm{fit2}}$ (keV)  & $\Gamma_{\textrm{fit3}}$ (keV)  & 
$\Gamma_{\textrm{fit4}}$ (keV)  & $\Gamma_{\textrm{fit5}}$ (keV)  \\ 
\hline \hline
$\rho^0 \rightarrow \eta \gamma$  & $44 \pm 3$ & $41 \pm 3$  & 41 $\pm$ 3 & 38 $\pm$ 2 & 38 $\pm$ 2 & 41 $\pm$ 3 \\
$\rho^0 \rightarrow \pi^0 \gamma$  & $69 \pm 9$ & $85 \pm 5$  & 85 $\pm$ 5 & 82 $\pm$ 2 & 79 $\pm$ 2 & 85 $\pm$ 5 \\
$\rho^+ \rightarrow \pi^+ \gamma$  & $67 \pm 7$ & $67 \pm 8 $  & 74 $\pm$ 2 & 75 $\pm$ 2 & 75 $\pm$ 2 & 74 $\pm$ 2 \\
$\omega \rightarrow \eta \gamma$  & $3.8 \pm 0.3$ & $4.0 \pm 0.5$  & 4.0 $\pm$ 0.5 & 3.4 $\pm$ 0.2 & 3.5 $\pm$ 0.2 & 4.0 $\pm$ 0.5 \\
$\omega \rightarrow \pi^0 \gamma$  & $713 \pm 20$ & $705 \pm 21$  & 701 $\pm$ 20 & 703 $\pm$ 19 & 704 $\pm$ 19 & 701 $\pm$ 20 \\
$\phi \rightarrow \eta \gamma$  & $55.4 \pm 1.1$ & $55 \pm 3$  & 55 $\pm$ 8 & 54 $\pm$ 8 & 54 $\pm$ 8 & 55 $\pm$ 3 \\
$\phi \rightarrow \eta^{\prime} \gamma$  & $0.26 \pm 0.01$ & $0.27 \pm 0.01$  & 0.27 $\pm$ 0.04 & 0.28 $\pm$ 0.05 & 0.27 $\pm$ 0.05 & 0.27 $\pm$ 0.01 \\
$\phi \rightarrow \pi^0 \gamma$  & $5.5 \pm 0.2$ & $5.5 \pm 1.0$  & 5.5 $\pm$ 1.1 & 5.5 $\pm$ 0.3 & 5.5 $\pm$ 0.3 & 5.5 $\pm$ 1.0 \\
$\eta^{\prime} \rightarrow \rho^0 \gamma$  & $57 \pm 3$ & $57 \pm 4$  & 57 $\pm$ 4 & 56 $\pm$ 3 & 55 $\pm$ 3 & 57 $\pm$ 4 \\
$\eta^{\prime} \rightarrow \omega \gamma$  & $5.1 \pm 0.3$ & $5.2 \pm 0.2$  & 5.2 $\pm$ 0.2 & 6.4 $\pm$ 0.1 & 6.5 $\pm$ 0.1 & 5.2 $\pm$ 0.2 \\
$K^{*0} \rightarrow K^0 \gamma$  & $116 \pm 10$ & $116 \pm 11$  & 116 $\pm$ 10 & 116 $\pm$ 10 & 116 $\pm$ 10 & - \\
$K^{*+} \rightarrow K^+ \gamma$  & $46 \pm 4$ & $46 \pm 5$  & 46 $\pm$ 5 & 46 $\pm$ 5 & 46 $\pm$ 5 & - \\
\hline 
$\chi_{\textrm{min}}^2/\textrm{d.o.f.}$ & - &2.3 & 1.9 & 4.4 & 4.8 & 1.9 \\
\hline \hline
\end{tabular}
}
\label{tb4}
\end{table}

It is worth highlighting that the biggest contribution to the residual $\chi_{\textrm{min}}^2$ in $\Gamma_{\textrm{fit1}}$, $\Gamma_{\textrm{fit2}}$ and $\Gamma_{\textrm{fit5}}$
consistently comes from the neutral $\rho^0 \rightarrow \pi^0\gamma$ decay. 
This might be related to the fact that the measurement associated to this decay channel has relatively small experimental uncertainty. 
However, it might also be pointing to limitations directly connected to the assumptions that have been taken in the phenomenological model presented in this letter, 
such as, for example, potential gluonic content of the mesonic wavefunctions or contributions to the mixing from gluonic annihilation channels.



\section{\label{conclusions}Conclusions}

The phenomenological model based on the standard quark model with two sources of flavour $SU(3)$-symmetry breaking proposed by Bramon et al.~in Ref.~\cite{Bramon:2000fr} 
has been tested using the most up-to-date $VP\gamma$ experimental data \cite{Tanabashi:2018oca} in section~\ref{sect3}. 
It has been shown that the quality of the most recent empirical data is sufficiently good to see that the model struggles to accurately reproduce experiment. 
Consequently, the objective of the present work has been to enhance this phenomenological model to reconcile it with experiment. 
This has been achieved by introducing isospin symmetry-breaking effects into the model.

The main result drawn from the present investigation is that the quality of the most up-to-date experimental data \cite{Tanabashi:2018oca} enables a small amount of isospin-symmetry breaking that is inconsistent with zero, with a confidence level of approximately $2.5\sigma$, 
using the enhanced phenomenological model. 
The quality of the performed fits is good, with e.g.~$\chi^2_{\textrm{min}}/\textrm{d.o.f.} \simeq 1.9$. 
In addition, the estimations for the fit parameters appear to be very robust across the fits that have been performed. 
The fitted values for $g = 0.69 \pm 0.01$ GeV$^{-1}$, $\phi_{23} = (41.5 \pm 0.5)^\circ$, $\phi_V = (4.0 \pm 0.2)^\circ$ and $m_s/\overline{m} = 1.17 \pm 0.06$ 
are in good agreement with those from other analysis available in the published literature (e.g.~\cite{Escribano:2005qq,Escribano:2007cd,Ambrosino:2009sc}). 
Contrary to this, our estimates for the parameters controlling the mixing in the $\pi^0$-$\eta$ and $\pi^0$-$\eta^{\prime}$ sectors, 
i.e.~$\epsilon_{12}=(2.4\pm1.0)\ \%$ and $\epsilon_{13}=(2.5\pm0.9)\ \%$ 
(using the mathematical definition from Eq.~(\ref{eq153})) or $\epsilon=\epsilon_{\pi\eta}=(0.1\pm0.9)\ \%$ and $\epsilon^{\prime}=\epsilon_{\pi\eta^{\prime}}=(3.5\pm0.9)\ \%$ 
(once translated into Kroll's \cite{Kroll:2005sd} and Escribano et al.'s \cite{Escribano:2016ntp} definitions), 
are not in accordance with the estimations that were provided by these authors in Ref.~\cite{Kroll:2005sd} and \cite{Escribano:2016ntp}.

To conclude, it is worth highlighting that all the results from the present investigation appear to indicate that a phenomenological model including simple quark model concepts, 
with a quantum mechanical extension implementing a second source of flavour symmetry breaking, is still sufficient to describe to a large degree of accuracy the radiative decays, 
and the rich and complex mixing phenomenology in the pseudoscalar meson sector.



\section*{Acknowledgements} 

The work of R.~Escribano is supported by the Secretaria d'Universitats i Recerca del Departament d'Empresa i Coneixement de la Generalitat de Catalunya 
under the grant 2017SGR1069, by the Ministerio de Econom\'{i}a, Industria y Competitividad under the grant FPA2017-86989-P, 
and from the Centro de Excelencia Severo Ochoa under the grant SEV-2016-0588. 
This project has received funding from the European Union's Horizon 2020 research and innovation programme under grant agreement No.~824093.





\end{document}